\documentclass[twocolumn,trackchanges,floatfix]{aastex631}
\usepackage{lineno}

\accepted{ in ApJL / 06-Aug-2025}

\shorttitle{UV-bright White Dwarf stars in NGC\,362}
\shortauthors{Yadav \& Dattatrey et al. (2025)}
\graphicspath{{./}{figures/}}

\begin{document}

\title{Detection of young massive white dwarfs in core-collapsed globular cluster NGC 362}
\author{R.K.S. Yadav}  
\affiliation{ Aryabhatta Research Institute of Observational Sciences, Manora Peak, Nainital 263002, India.} 

\author{Arvind K. Dattatrey}
\affiliation{ Aryabhatta Research Institute of Observational Sciences, Manora Peak, Nainital 263002, India.}
\affiliation{ Deen Dayal Upadhyay Gorakhpur University, Gorakhpur, Uttar Pradesh 273009, India.} 

\author{Annapurni Subramaniam}
\affiliation{Indian Institute of Astrophysics, Koramangala, Bangalore 560034, India.}

\author{Geeta Rangwal}
\affiliation{ Aryabhatta Research Institute of Observational Sciences, Manora Peak, Nainital 263002, India.}

\author{Ravi S. Singh}
\affiliation{Maa Pateshwari University, Balarampur -271201 (U.P), India}
\affiliation{ Deen Dayal Upadhyay Gorakhpur University, Gorakhpur, Uttar Pradesh 273009, India.} 

\email{rkant@aries.res.in}
\email{Corresponding author: Physics.arvind97@gmail.com}


\begin{abstract}
Core-collapsed globular clusters are ideal targets to explore the presence of stellar collision products. Here, we have studied seventeen FUV bright white dwarf members in the globular cluster NGC 362 using data obtained from the Ultra Violet Imaging Telescope (UVIT) mounted on AstroSat and from HST. Multi-wavelength spectral energy distributions (SEDs) are analyzed using UV and optical data sets to characterize and determine the parameters of white dwarfs. Fourteen of the white dwarfs fit single-component SEDs well, while three showed a good fit with a two-component SED model, indicating a binary system comprising a white dwarf and a low-mass main-sequence star. The effective temperature, radius, luminosity, and mass of white dwarfs range between 22000 - 70000  K, 0.008 - 0.028 R$_\odot$, 0.09 - 3.0 L$_\odot$, and 0.30 - 1.13 M$_\odot$, respectively. The effective temperature, radius, luminosity, and mass of the companions (low-mass main-sequence stars) are 3500 - 3750  K, 0.150 - 0.234 R$_\odot$, 0.003 - 0.01 L$_\odot$, and 0.14 - 0.24 M$_\odot$, respectively. The three binary systems (WD-MS), along with the massive WDs may have formed through dynamical processes that occurred during the core collapse of the cluster. This is the first evidence of a massive WD formation in a core-collapsed cluster, which is the missing link in the formation of a fast radio burst (FRB) progenitor in a globular cluster. This study provides evidence that NGC 362 hosts stellar systems that may evolve into exotic stars such as Type Ia supernovae, and/or FRBs in the future. *This is paper VI of the Globular Cluster UVIT Legacy Survey. 
\end{abstract}



\section{Introduction}
\label{sec:intro}

White dwarfs (WDs) are the remains of low to intermediate-mass main sequence stars \citep{2010MmSAI..81..908A,2022yCat..75185106J}. Globular clusters are expected to contain tens of thousands of WDs. However, the observational study of these WDs has been slow due to their low luminosity \citep{1978ApJ...224L...9R, 1995AJ....110..682E}. In the mid-1990s, significant progress was made with the observation of large WD populations and the WD cooling sequence in various globular clusters \citep{1995ApJ...451L..17R,1997ApJ...484..741R,1996ApJ...468..655C, Renzini_1996, 2007ApJ...671..380H, Calamida_2008}. 

Additionally, the study of white dwarf-main sequence (WD-MS) systems is an important research area because they constitute a common final stage object of stellar evolution, a WD, and the most frequent type of star: the M spectral type component. Close WD-MS binaries are also progenitor candidates for cataclysmic variables (CVs) and Type Ia supernovae \citep{refId0, 2019A&A...622..A35L}.

The N-body simulations of globular clusters have shown that massive WDs can be found in the central regions of core-collapsed GCs \citep{2020ApJS..247...48K,2021ApJ...912..102R}. 
Recent studies show that NGC 6397 harbours a massive central region with the possibility of hosting a few stellar black-holes at present day \citep{2020ApJ...898..162W,2021ApJ...912..102R,2022MNRAS.514..806V}. \cite{2009ApJ...699...40S} found that some of the He WDs might be binaries with massive dark CO WD companions.
\cite{2025ApJ...979..167P} detected WDs in the mass range 0.18 M$_\odot$ $\le$ M $\le$ 0.5 M$_\odot$ and a second candidate for a magnetic He core WD. NGC 6397 is, therefore, a core-collapsed globular cluster \citep{1995AJ....109..218T} that harbours exotic systems, and \cite{Kremer_2021} used this cluster to simulate WD systems in such systems.

The recent discovery of a Fast Radio Burst (FRB) repeater in an old globular cluster in M81 \citep{2022Natur.602..585K} has prompted the astronomical community to investigate potential progenitors. The magnetar model has emerged as the most accepted model for FRBs \citep{Petroff_2019}. There have been studies to explore and identify probable progenitor populations for an FRB in a globular cluster \citep{2023ApJ...944....6K}.  A massive white dwarf binary merger, as suggested by \cite{Kremer_2021}, may provide a natural formation mechanism for this repeater. 

Recently, \cite{2023MNRAS.523L..58D} detected WD-MS binaries, apart from the detection of WD companions to blue stragglers in NGC 362 \citep{2023ApJ...943..130D}. They attributed many binary systems to the recent core-collapse of the cluster \citep{2023MNRAS.523L..58D,2023ApJ...943..130D}. As this is a core-collapsed cluster, it is important to check its WD population. This study aims to detect and analyze the white dwarf systems in the globular cluster NGC 362 to determine their fundamental parameters and nature.  

\begin{figure*}
\centering
\includegraphics[width=0.88\textwidth]{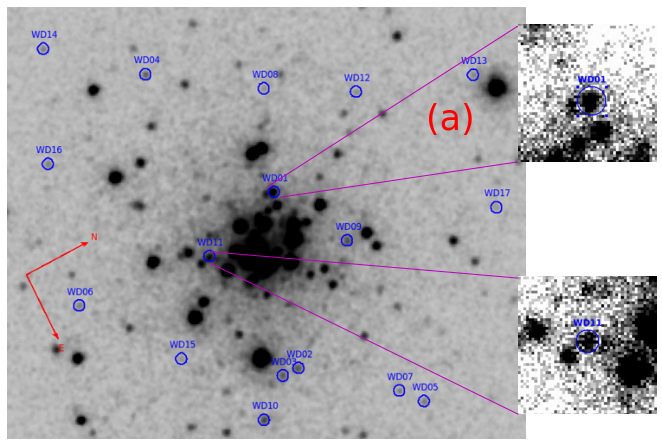} 
\includegraphics[width=0.345\textwidth]{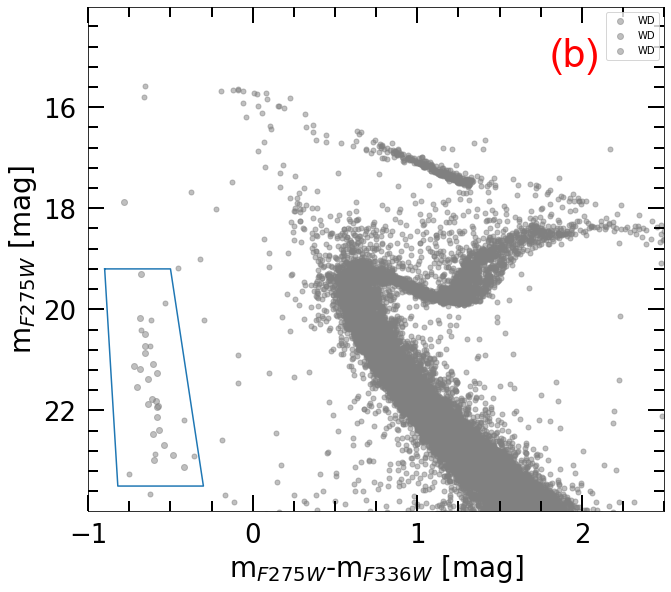} 
\centering
\includegraphics[width=0.32\textwidth]{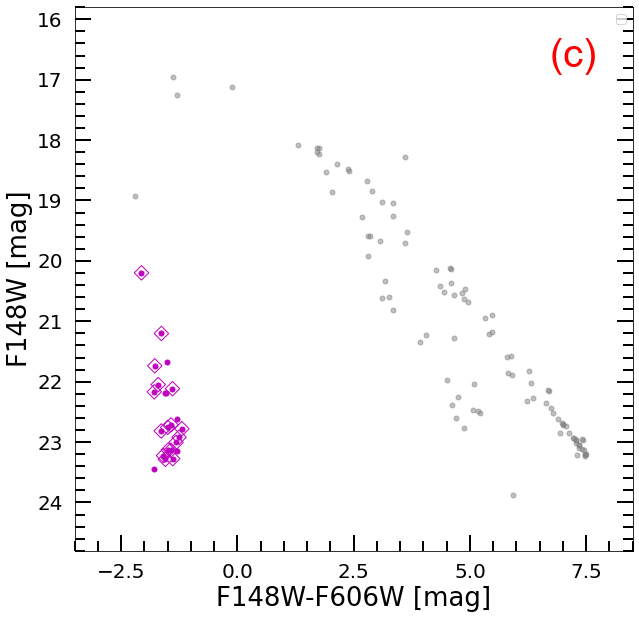}
\includegraphics[width=0.32\textwidth]{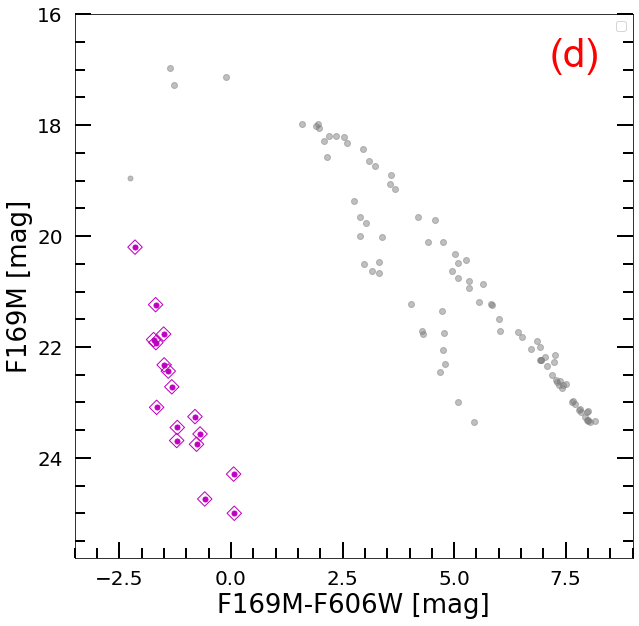}

\caption{Panel (a): The FUV image of 1.75$^{\prime}\times 1.75^{\prime}$ for the cluster NGC 362 in the F148 filter. North is up, while the east is on the left. The seventeen FUV bright white dwarfs are shown with the star ID numbers. 
The panel (b) shows the (m$_{F275W}$-m$_{F336W}$) vs m$_{F275W}$ while the (c) and (d) panel shows the (F148W-F606W) vs F148W and (F169M-F606W) vs F169M CMDs. A magenta color in panels (c) and (d) represents the WDs. The open magenta squares represent the WDs without contamination from neighbouring stars.}
\label{uv_cmd}
\end{figure*}

The paper starts with an introduction in Section \ref{sec:intro}, followed by a description of observations and methodology in Section \ref{observation} and \ref{selection}. Section \ref{SED} provides details about the SEDs and the parameters of the WDs, with the H-R diagram of WDs and their companions in Section \ref{hrdc}, followed by discussions in Section \ref{dis}. A summary and conclusions of our work are given in the final section.

\begin{figure*}
\centering
\includegraphics[width=18cm, height=4.5cm]{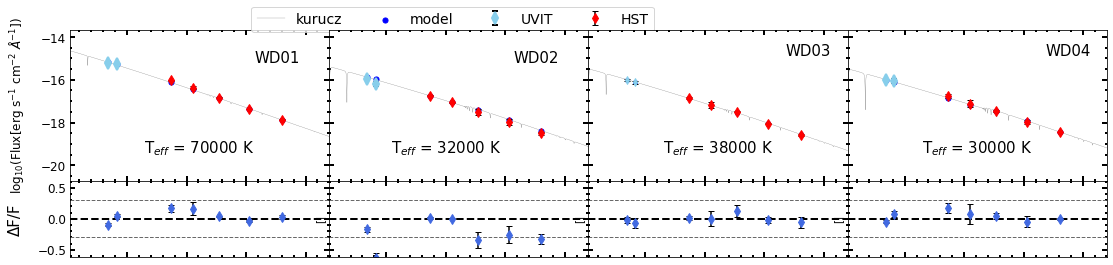}
\includegraphics[width=18cm, height=4.5cm]{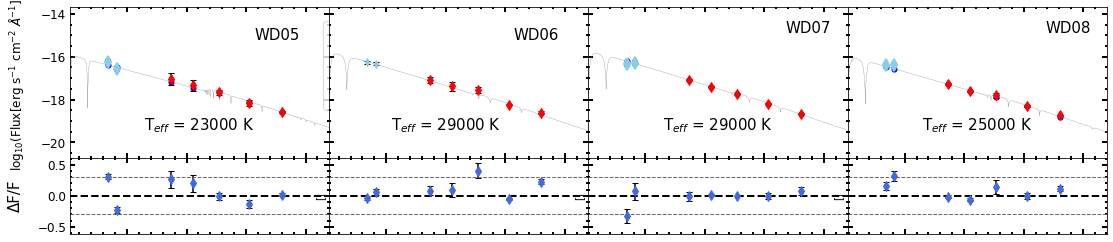}
\includegraphics[width=18cm, height=4.5cm]{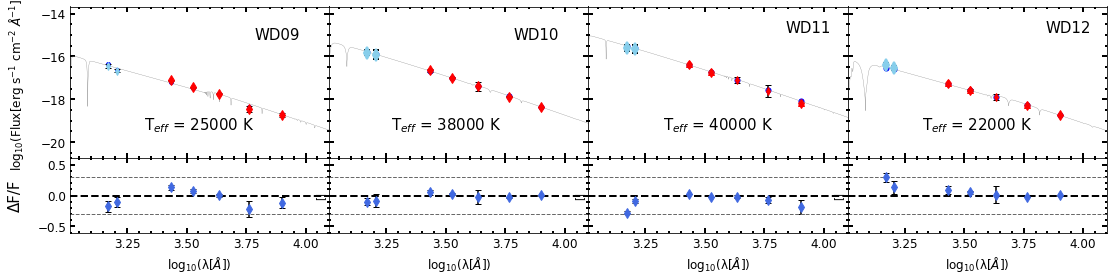}
\includegraphics[width=16cm, height=4.5cm]{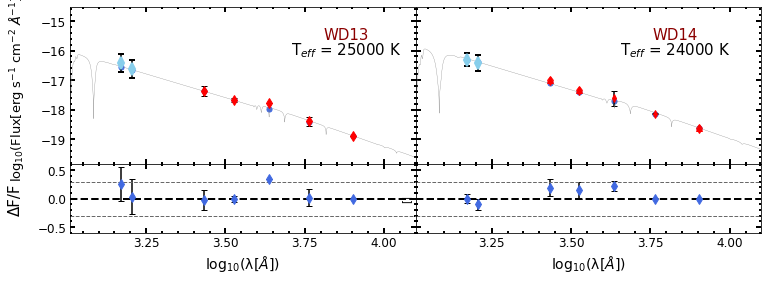}
\includegraphics[width=18cm, height=4.81cm]{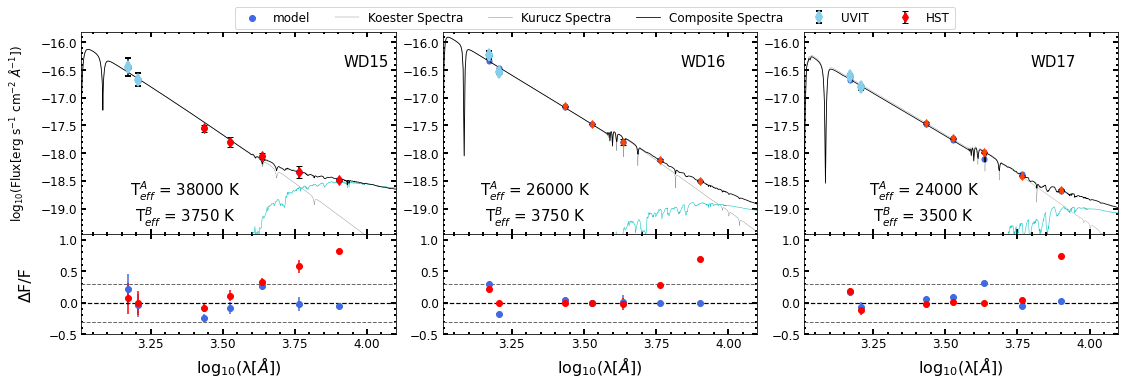}
\caption{The SEDs of the fourteen single-component and three two-component (WD15, WD16 and WD17) WDs. The blue points represent the synthetic flux from the model used to fit the observed SED. The T$_{eff}$ of the hot (A) and cool (B) components are displayed for the two-component WDs.  The UVIT and HST data points are represented by cyan and red color, while the model data is represented with blue points. The \cite{2010MmSAI..81..921K}, \cite{2003IAUS..210P.A20C}, and combined spectra are shown by grey, cyan, and black curves, respectively. $\Delta F/F$ is the fractional residual. }
\label{sed}
\end{figure*}

\section{Observations} 
\label{observation}
 The observational data for the globular cluster NGC 362 in UV were collected using an ultraviolet imaging Telescope (UVIT) mounted on Astrosat, India's first multi-wavelength space observatory. The observations took place on November 11, 2016, using two UV filters: F148W and F169M. The exposure times for the filters were 4900 and 4600 seconds, respectively. The identification map for the cluster in filter F148 is depicted in Figure \ref{uv_cmd}(a). The raw images were processed using CCDLAB software \citep{2017PASP..129k5002P}, which corrects for satellite drift, flat-field, geometric distortion, fixed pattern noise, and cosmic rays. Detailed descriptions of the telescope, instruments, and preliminary calibration can be found in \cite{2016ApJ...833L..27S} and \cite{2017AJ....154..128T}. The photometry was done using the DAOPHOT routine developed by \cite{1987PASP...99..191S}. We conducted Point Spread Function (PSF) photometry, a technique particularly useful for analysing crowded regions. To determine the PSF, we used several well-isolated brighter stars. By analyzing the light distribution of isolated stars, we generated a model of the PSF for a given image. The PSF model is then fitted to each detected star in the image. Once the PSF is fitted, it is subtracted from the star's image, effectively removing the contribution of that star's light from the surrounding pixels. As a result, we obtained PSF magnitudes for all the stars in the image. The photometric depth in the F148 band is $\sim$ 23.0 mag.\\

\section {Selection of white dwarfs} 
\label{selection}

We used the NUV-FUV and FUV-optical CMDs to select the WDs for the present analysis. We compared our UVIT catalogue with the HST UV HUGS catalogue \citep{2018MNRAS.481.3382N} to find the optical counterparts of WDs. The HUGS catalogue provides the HST magnitudes in F275, F336, F438, F606, and F814 filters. In Figure \ref{uv_cmd}(b), we have plotted the HST CMD  in the (m$_{F275W}$-m$_{F336W}$), m$_{F275W}$ plane.

This diagram shows a hot stellar population, including horizontal branch stars, blue stragglers, and white dwarfs. 
There is a vertical sequence ranging from $\sim$ 19 to 23 in m$_{F275W}$ mag and with a color of approximately -0.65 mag in (m$_{F275W}$-m$_{F336W}$), representing the WD sequence. A blue trapezoidal box depicts the boundary of the WD sequence. In this way, we identified 33 WDs in the blue rectangle box.

In Figure \ref{uv_cmd}(c) and \ref{uv_cmd}(d), we present the FUV-optical CMD ((F148-F606) vs F148 and (F169M-F606) vs F169M) of the hot population of the cluster. We found 23 FUV bright WDs common in the HST and UVIT catalogue and shown with magenta points. Furthermore, we checked the contamination in flux from nearby stars of these WDs by comparing the aperture magnitudes with the PSF magnitudes. The PSF photometry reduces contamination by $\sim$3\% if any contamination is present. Finally, we found 17 non-contaminated WDs having UV and optical data and marked them in Figure \ref{uv_cmd}(a) with their IDs. All 17 WDs are members of the cluster based on the analysis by \cite{2017AJ....153...19S}. We considered these WDs for further study. 

\section {Spectral energy distribution of white dwarfs} \label{SED}

In this section, we build the SED for 17 WDs to analyze their basic parameters and characteristics using the virtual observatory tool, VOSA (VO SED Analyzer \citep{2008A&A...492..277B}). The reddening $E(B-V)=0.05$ mag from \cite{2010arXiv1012.3224H} and distance of 8.83 Kpc are taken from \citep{2021MNRAS.505.5978V}. VOSA utilizes filter transmission curves to calculate synthetic photometry based on a selected theoretical model. The synthetic photometric data is then compared to the observed data, and the best fit is determined using the $\chi^2_{r}$ minimization test, as explained in \cite{2023ApJ...943..130D}. In summary, we fitted the \cite{2010MmSAI..81..921K} model, considering variations in temperature and log $g$. The effective temperatures ranged from 5000 - 80000 K, with log $g$ ranging from 5 - 9.5 dex. The SED was generated using photometric data covering UV to optical wavelengths. Initially, we fitted a SED with a single component, as shown by the grey curve in Figure \ref{sed}. The data points used for fitting are represented by cyan and red-filled circles with error bars, while model data points are shown in blue. The difference between the fitted model and the observed fluxes, normalized by the observed flux, is displayed in the lower panel of Figure \ref{sed}. The dashed horizontal lines at $\pm$0.3 (30\%) represent the threshold. Except for the WDs WD15, WD16, and WD17, the residual plot for all the WDs is within $\pm$0.3. This suggests that a single SED fits all WDs except WD15, WD16, and WD17. The derived parameters and the fitting parameters for the single-component WDs are listed in Table \ref{para}. 

The residual plot for the WDs WD15, WD16, and WD17 reveals that the near-IR residual flux is more significant than 0.3, as shown with the red points in the bottom panels of Figure \ref{sed}. The residuals indicate that a single-component SED does not fit well in the near-IR region. The fitting shows a near-IR excess in the SEDs. Therefore, a combination of hot and cool spectra is tried. A two-component SED is fitted for the WDs WD15, WD16, and WD17, as depicted in Figure \ref{sed}. The grey curve represents the \cite{2010MmSAI..81..921K} model corresponding to the hot component, while the \cite{2003IAUS..210P.A20C} model shown in the sky blue curve represents the cool component. The combined spectrum is shown with a black curve. The residual plot represented with blue points shows that the combined spectrum fits well. The fitting parameters, along with the characteristic parameters, are listed in Table \ref{para}.

The fundamental parameters of the 17 WDs derived using SED fit are listed in Table \ref{para}. For WDs, the fundamental parameters are in the range of T$_{eff}$ $\sim$ 22000-70000 K, log $g$ $\sim$ 6.5 – 9.5, $L$ $\sim$ 0.09 – 3.0 $L_{\odot}$, and $R$ $\sim$ 0.008 – 0.028 $R_{\odot}$. The parameters of the cool companions of WD15, WD16 and WD17 are in the range T$_{eff}$ $\sim$ 3500-3750 K, log $g$ =0.5-3.0, [Fe/H] $\sim$ -1.0, $L$ $\sim$ 0.003-0.010 $L_{\odot}$ and $R$ $\sim$ 0.150-0.234 $R_{\odot}$. The goodness of fit parameters, i.e., $\chi^2_{r}$, $V_{gf}$, and $V_{gfb}$, are also listed in Table \ref{para}. The $V_{gf}$ and  $V_{gfb}$ are the modified reduced $\chi^2_{r}$ provided by VOSA by assuming the observational errors of 2\% and 10\% of the observed flux, respectively. These parameters provide a visual check of the fit quality, with recommended thresholds of $V_{gf}$ $<$ 25 or  $V_{gfb}$ $<$ 15 \citep{2018MNRAS.480.4505J, 2021MNRAS.506.5201R}.

\begin{table*}

\caption{The best-fit parameters for the WDs and their companions. The positions (RA and DEC) are provided in degrees, and $r$ (in arcsec) denotes the distance from the cluster center. The $T_{eff}$ is the effective temperature in K, log g is the surface gravity in the logarithm unit, the reduced chi-square ($\chi^2_{r}$),  the luminosity ($L$) and radius ($R$) are in Solar unit while $V_{gf}$ and	$V_{gfb}$ are the visual goodness of fit. The $N_{fit}$ is the number of points considered in the fitting, and $N_{tot}$ is the total number of points. $M$ is the mass of the stars in Solar units.}

\resizebox{18cm}{!}{%

\begin{tabular}{
    p{1.4cm} p{1.5cm} p{1.65cm} p{0.65cm} 
    p{2.1cm} p{1.2cm} p{1.8cm} p{1.8cm} 
    p{1.2cm} p{1cm} p{0.5cm} p{1.2cm} 
    p{1.5cm} p{1.8cm} p{1.1cm}
}
\hline
Name & RA & DEC & $r$ & $T_{\text{eff}}$ (K) & $\log g$ & $L/L_{\odot}$ & $R/R_{\odot}$ & $\chi^2_r$ & $V_{\text{gf}}$ & $V_{\text{gfb}}$ & $N_{\text{fit}}/N_{\text{tot}}$ & $M/M_{\odot}$ & Age (Myr) \\
\hline
WD01     &    15.79445  &    $-$70.84506 & 20.3	& 70000$^{+5000}_{-10000}$ &  6.5 &    2.997$^{+0.160}_{-0.180}$ & 0.028$^{+0.002}_{-0.003}$ &   31.48&     5.82& 	 1.15& 	 7/7 &0.86$^{+0.08}_{ -0.16}$ &0.7 \\ 
WD02     &    15.84468   &    $-$70.85072 & 40.0	& 32000$^{+2000}_{-500}$   &  6.5 &    0.434$^{+0.010}_{-0.020}$ & 0.022$^{+0.001}_{-0.002}$ &   43.12&     8.62& 	 4.33& 	 7/7 &0.41$^{+0.12}_{-0.04}$ &0.6\\ 
WD03     &    15.84460    &    $-$70.85229 & 45.3	& 38000$^{+2000}_{-2000}$  &  6.5 &    0.415$^{+0.010}_{-0.020}$ & 0.017$^{+0.001}_{-0.001}$ &   36.83&     0.74& 	 0.49& 	 7/7 &0.62$^{+0.07}_{-0.06}$ &1.5\\ 
WD04     &    15.74530    &    $-$70.85113 & 75.4	& 30000$^{+500}_{-1000}$   &  6.5 &    0.405$^{+0.026}_{-0.040}$ & 0.024$^{+0.001}_{-0.001}$ &    2.97&     0.59& 	 0.44& 	 7/7 &0.35$^{+0.02}_{-0.03}$ &0.6\\ 
WD05     &    15.87124   &    $-$70.84074 & 79.0	& 23000$^{+500}_{-1000}$   &  6.5 &    0.166$^{+0.011}_{-0.019}$ & 0.025$^{+0.001}_{-0.001}$ &   23.17&     4.63& 	 3.43& 	 7/7 &0.30$^{+0.02}_{-0.10}$ &2.2\\ 
WD06     &    15.79724   &    $-$70.86747 & 69.4	& 29000$^{+3000}_{-1000}$  &  9.5 &    0.185$^{+0.040}_{-0.010}$ & 0.017$^{+0.001}_{-0.002}$ &   25.84&     5.17& 	 2.90& 	 7/7 &0.50$^{+0.14}_{-0.03}$ &8.0\\ 

WD07    &     15.86496   &    $-$70.84249 & 72.2	& 29000$^{+1000}_{-2000}$  &  9.5 &    0.187$^{+0.010}_{-0.010}$ & 0.018$^{+0.002}_{-0.001}$ &   23.19&     4.46& 	 2.32& 	 7/7 &0.50$^{+0.04}_{-0.06}$ &8.0\\ 
WD08    &     15.76577   &    $-$70.84111 & 56.8	& 25000$^{+500}_{-1000}$   &  8.5 &    0.125$^{+0.001}_{-0.010}$ & 0.018$^{+0.001}_{-0.001}$ &   31.92&     6.38& 	 3.18& 	 7/7 &0.43$^{+0.02}_{-0.04}$ &10\\ 
WD09    &     15.79257   &    $-$70.85373 & 26.0	& 25000$^{+1000}_{-1000}$  &  6.5 &    0.139$^{+0.010}_{-0.010}$ & 0.020$^{+0.001}_{-0.002}$ &   58.59&     11.7& 	 7.69& 	 7/7 &0.41$^{+0.04}_{-0.04}$ &5.0\\  
WD10    &  15.854263    &    $-$70.856049 & 61.3    & 38000$^{+2000}_{-2000}$  &  9.5 &    0.567$^{+0.040}_{-0.020}$ & 0.021$^{+0.007}_{-0.005}$ &	 31.06&	     3.2&	 0.436&   7/7&0.54$^{+0.05}_{-0.05}$ &3.5	\\
WD11	&  15.80346	    &    $-$70.853723 & 19.4    & 40000$^{+5000}_{-2000}$  &  6.5 &    1.222$^{+0.050}_{-0.060}$ & 0.028$^{+0.006}_{-0.005}$ &	 71.20&	    7.98&	 1.92& 	  7/7&0.41$^{+0.10}_{-0.03}$ &0.2\\
WD12	&  15.78068	    &    $-$70.833015 & 65.2    & 22000$^{+2000}_{-1000}$  &  9.5 &    0.098$^{+0.04}_{-0.003}$  & 0.020$^{+0.010}_{-0.008}$ &	 46.83&	     5.1 &   2.11&	  7/7&0.35$^{+0.07}_{-0.03}$ &7.0\\
WD13	&  15.792786	&    $-$70.821661 & 97.5    & 25000$^{+1000}_{-500}$   &  8.5 &    0.096$^{+0.017}_{-0.003}$ & 0.016$^{+0.010}_{-0.004}$ &	 07.07&	     1.76&	 1.65&	  6/7&0.50$^{+0.05}_{-0.02}$ &15\\
WD14	&  15.724402	&    $-$70.859252 & 105.2   & 24000$^{+1000}_{-1000}$  &  9.5 &    0.155$^{+0.010}_{-0.010}$ & 0.022$^{+0.010}_{-0.007}$ &	 08.65&	     2.88&	 1.99&	  6/7&0.35$^{+0.04}_{-0.03}$ &7.7\\
\hline
WD15A 	&     15.82578  &    $-$70.86067 & 49.0     & 38000$^{+1000}_{-1000}$  &  9.5 &    0.096$^{+0.005}_{-0.001}$ & 0.008$^{+0.001}_{-0.002}$ &   67.35&     35.5& 	 12.2& 	  7/7&1.13$^{+0.02}_{-0.02}$ &35\\ 
WD15B   &               &                &          &  3750$^{+250}_{-250}$    &  0.5 &    0.010$^{+0.001}_{-0.001}$& 0.234$^{+0.016}_{-0.032}$  &        &         & 	     & 	    &0.24$^{+0.01}_{-0.01}$&                          \\ 

WD16A  &    15.755527	&    $-$70.863951 & 83.2    & 26000$^{+500}_{-1000}$   &  7.0 &	   0.177$^{+0.003}_{-0.016}$& 0.0202$^{+0.007}_{-0.001}$&   71.69 &	     32.3&	  4.67&	  7/7&0.39$^{+0.02}_{-0.03}$ &5.2	\\
WD16B  &                &                &          &  3750$^{+125}_{-250}$    &  3.0 &	   0.004$^{+0.001}_{-0.0007}$&0.1503$^{+0.001}_{-0.001}$&	             &     &		&   &0.16$^{+0.01}_{-0.01}$&\\	

WD17A &   	15.830937	&   $-$70.825606 & 86.2& 24000$^{+500}_{-4000}$     &	7.5	 &     0.087$^{+0.004}_{-0.005}$&	0.0162$^{+0.007}_{-0.001}$&	8.480&  8.48&	7.75&	7/7	&0.48$^{+0.03}_{-0.14}$&15\\
WD17B  &                &                & &  3500$^{+125}_{-125}$	  & 3.0 &	   0.003$^{+0.001}_{-0.002}$&	0.154$^{+0.010}_{-0.010}$&	     &      &       &  &0.14$^{+0.01}_{-0.01}$&\\

\hline
\end{tabular}}
\label{para}
\end{table*}

\begin{figure}
\centering
\includegraphics[width=8.0cm, height=7.5cm]{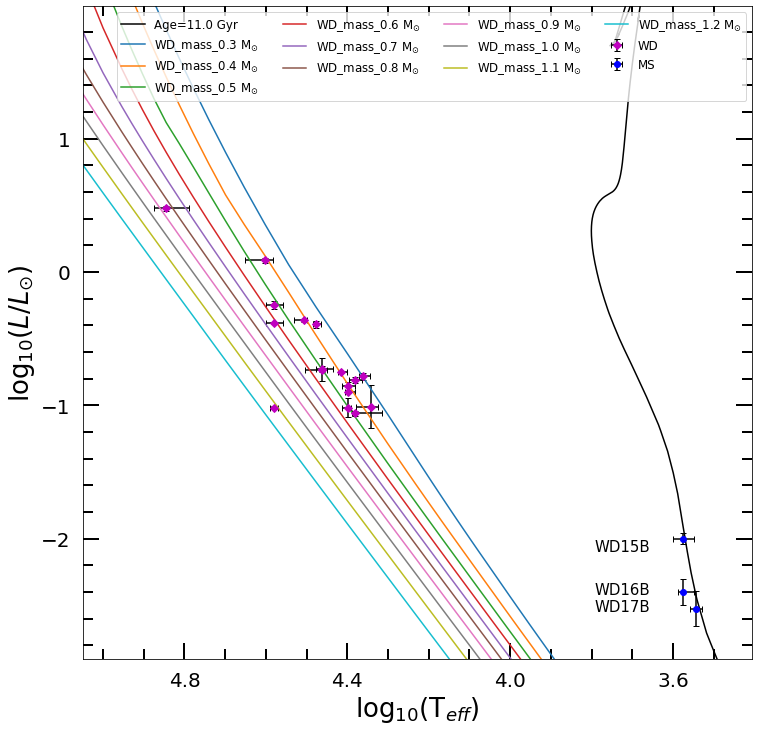}
\caption{The H-R diagram of hot and cool components of the WDs. The red points represent the WDs, while the blue points represent the cool components (MS stars). The WDs cooling curves for different masses are shown with colors. The black curve represents the isochrone taken from  \citet{2018ApJ...856..125H} and \citet{2021ApJ...908..102P}.}
\label{hrd}
\end{figure}

\section{H-R diagram of the White Dwarfs and their companions}
\label{hrdc}

After obtaining the essential parameters of WDs, we created the HR diagram in Figure \ref{hrd}. The WDs are shown with magenta points, while the blue point represents the cool companions of WD15, WD16, and WD17.  The theoretical isochrone \citep{2018ApJ...856..125H,2021ApJ...908..102P} is plotted with the black line using the fundamental parameters of the cluster taken from \cite{2010arXiv1012.3224H}. We superimposed WD cooling curves taken from \cite{2009ApJ...696.1755T} for various masses, distinguished by colors. Our analysis utilized the DA models of WDs.  We are unable to clearly distinguish between DA and DB WDs based on the current analysis. This limitation in differentiating between the two types of WDs has led us to apply a DA model for our analysis. To determine the mass of the white dwarfs, we performed interpolation using their cooling curves. We generated 10 sub-models based on theoretical cooling curves, each within a mass interval of 0.1 M$_{\odot}$, to improve the accuracy of our mass estimates. In this way, we estimated the white dwarf masses from 0.30 to 1.13 M$_{\odot}$, with a cooling age ranging from 0.2 to 35 million years.  We also assessed the mass and age of the WDs using DB models, and the results were consistent within the margin of error. We note that the WDs were formed very recently and have a large mass range.

{\bf Mass of the cool companions:} The cool companions (WD15B, WD16B, WD17B) follow the isochrone of 11 Gyr and are located in the lower part of the main sequence. We estimated the mass of cool companions in the range 0.14 - 0.24 $M_{\odot}$. Based on the location in the H-R diagram and parameters derived in the previous section for the cool companions, we may conclude that they may be  M-type dwarf stars on the MS. 

\label{dis}

{\bf Low mass WDs:} The minimum mass of WDs formed through single star evolution is 0.4 $M_{\odot}$ \citep{2010ApJ...723.1072B}. Out of the seventeen WDs studied, three (WD04, WD05, and WD14) fall within the mass range of 0.3 - 0.4 $M_{\odot}$. According to the \citet{2007ApJ...671..761K}, the low-mass helium-core white dwarfs ($M <$ 0.45 $M_{\odot}$) can be produced from interacting binary systems, and generally, all of them have been attributed to this channel. However, these WDs may also arise from a single star that experiences significant mass loss during the red giant branch phase. 
 
The single low-mass WDs are detected in many studies. A PG survey done by \cite{2005yCat..21560047L} found 30 low-mass WDs. About 47\% of the PG low-mass WDs searched for companions seem single \citep{2007ApJ...671..761K}. The ESO SN Ia Progenitor Survey searched for radial velocity variations in more than a thousand WDs using the Very Large Telescope \citep{2001AGM....18S0910N}. They have found that 15 of these low-mass WDs do not show any radial velocity variations, corresponding to a single low-mass WD \citep{2007ASPC..372..387N}. 

{\bf Normal and slightly massive WDs:} Additionally, twelve WDs (WD02, WD03, WD06, WD07, WD08, WD09, WD10, WD11, WD12, WD13, WD16, and WD17) fall within the mass range of 0.4 - 0.6 $M_{\odot}$. Among these,  WD16 and WD17 are binary systems, with WD masses of 0.39$M_{\odot}$ and 0.48 $M_{\odot}$, and companion masses of 0.16 $M_{\odot}$ and 0.14 $M_{\odot}$, respectively. The WDs having a mass between 0.4 - 0.6 $M_{\odot}$ are likely to have formed via single-star evolution. In globular clusters, WDs are currently produced by progenitors with stellar masses of about 0.8-1.0 $M_{\odot}$ \citep{moehler2011whitedwarfsglobularclusters}. The cooling age of these WDs is 0.2 to 15 Myr, indicating that they have formed quite recently.  As the progenitor mass is slightly more than the turn-off mass, some of the progenitor stars may be blue stragglers, which could have formed as a result of stellar mergers or mass transfer between stars in binary systems. Over time, blue stragglers evolve and, eventually, end their lives as white dwarfs \citep{moehler2011whitedwarfsglobularclusters, Ferraro_2009}. \cite{2023ApJ...943..130D}  detected extremely low-mass WD companions to 12 blue stragglers in this cluster. They also found that the blue stragglers of this cluster may have a mass up to 1.64 $M_\odot$. The massive blue stragglers are likely to be the progenitors of WDs with mass between 0.5 - 0.6 $M_{\odot}$. We therefore note that this cluster contains both blue stragglers and their progeny. 

{\bf WD-MS binaries:} In this analysis, we identified three binary systems (WD15, WD16 and WD17) composed of WD and MS stars. These systems may not be primordial binaries, as primordial binaries can be disrupted by dynamic processes within the cluster \citep{10.1111/j.1365-2966.2006.10876.x}. A simulation conducted by \cite{Kremer_2021} showed that 65\% WD-MS binaries form through dynamical interactions rather than as primordial binaries. \cite{2023MNRAS.523L..58D} found 4 WD-MS systems, where the MS stars were found to have low-mass or extremely low-mass WD companions. The above systems are likely to be mass-transfer products, whereas the WD-MS systems detected in this study may or may not have undergone mass-transfer. Indeed, a binary consisting of an MS star and a WD can form through several types of dynamical encounters: exchange interaction, tidal capture of an MS by a WD, or physical collisions between a red giant and an MS star \citep{10.1111/j.1365-2966.2006.10876.x}. Additionally, \cite{Kremer_2021} demonstrated that WD-MS binaries can also form through dynamical interactions in the core collapse globular clusters. Based on their cooling ages, we infer that these systems may have formed during the core-collapse phase of the cluster through dynamic processes. The typical mass of a WD capable of capturing an MS star is $\sim1.0\pm0.2$ 
 $M_{\odot}$ \citep{10.1111/j.1365-2966.2006.10876.x}. Numerical simulation conducted by \cite{1987ApJ...318..760S}
 indicate that within the core of a globular cluster, a 0.8 $M_{\odot}$ WD can collide with main-sequence stars of masses between 0.2 to 0.4 $M_{\odot}$. One of the WD-MS systems (WD15) has an ultra-massive WD (1.13 M$_\odot$) that may have formed via the collision of a WD binary, and it captures an MS star of mass 0.24 M$_\odot$ is therefore possible. A fraction of these stars can become bound to the WD, while the remainder will be dispersed. A portion of these dynamically formed WD-MS binary systems will initiate mass transfer and evolve into cataclysmic variables. 

{\bf Massive WDs:} The mass distribution of WDs shows a main peak at $\sim$ 0.6 $M_{\odot}$ and a smaller peak at the tail of the distribution $\sim$ 0.8 $M_{\odot}$ \citep{Kleinman_2013}. The presence of massive WDs ($M \ge$ 0.8 $M_{\odot}$) and ultra-massive WDs ($M \ge$ 1.10 $M_{\odot}$) have been discussed in several studies \citep{Hermes_2013, Kepler_2016, 10.1093/mnras/stx320}. In the present study, two WDs, WD1 and WD15, have masses greater than 0.8 $M_{\odot}$. The WD1 is classified as a massive WD, while WD15 is categorized as an ultra-massive WD.

Based on velocity dispersion analysis, \cite{Cheng_2020} concluded that about 20\% of the single WDs with masses greater than 0.8 $M_{\odot}$ form through the merger of two WDs. In another study, \cite{2020A&A...636A..31T} estimated, based on binary population synthesis studies, that from 30 to 45 \% of all the single WDs with masses larger than 0.9 $M_{\odot}$ are likely formed through binary mergers, and most of them via double WD mergers. Therefore, we expect that WD1 and WD15 white dwarfs may have formed due to the merger of binary WDs. 

\cite{2021NatAs...5.1170C} have studied the brightest portion of the WD cooling sequence in two twin old and massive globular clusters, M3 and M13. They found that the bright WDs in M13 may result from the cooling processes slowing down due to stable thermonuclear burning in the remaining hydrogen-rich envelope. Additionally, they concluded that this phenomenon occurs in M13 but not in M3, which is consistent with the different horizontal branch (HB) morphologies of the two clusters. The HB morphology of cluster NGC 362 resembles that of cluster M3, where such WDs are absent. Therefore, based on the HB morphology, we would not expect to find slow-cooling bright WDs in NGC 362.

\section{Discussion} 
\label{dis}
According to \cite{Dalessandro_2013} and \cite{2018ApJ...861...99L}, NGC 362 is a post-core-collapsed cluster, as indicated by its radial density profile and internal velocity dispersion. This cluster, therefore, is likely to show the signatures of core-collapse and is an ideal target to look for massive WDs. The WDs analyzed in this study have a range in mass and are very young (cooling age $\le$ 35 million years), suggesting that they have formed recently. Based on the cooling age, we infer that the massive WD (WD1) and the ultra-massive WD (WD15) may have formed during the core-collapse phase. During this phase, it is possible that binary WDs merged, resulting in the formation of massive WDs \citep{2020A&A...636A..31T, Kremer_2021}. A slow ($\sim$ 100 ms) radio pulsar was discovered in the globular cluster NGC 362 as part of the MeerKAT, TRAPUM project\footnote{https://www3.mpifr-bonn.mpg.de/staff/pfreire/GCpsr.html}. \cite{2023MNRAS.525L..22K} have suggested that slow pulsars in globular clusters may form relatively recently through the merging of WDs.  If this radio pulsar is indeed young, as has not yet been proven, we can expect that WD mergers are currently occurring in this cluster.  \cite{Kremer_2021}  studied WD systems in core-collapsed globular clusters and predicted the existence of young massive WDs that are similar to WD1 and WD15. We suggest that NGC 362 is a potential target to study core-collapse dynamics resulting in the formation of massive compact objects.

The massive WDs, WD01 and WD15 are located at distances of 20$^{\prime\prime}$ and 49$^{\prime\prime}$ from the center of the cluster. While they are outside the core radius of 11$^{\prime\prime}$, they fall within the half-light radius of 49$^{\prime\prime}$.2 of the cluster. According to the simulation by \cite{Kremer_2021}, massive WDs are expected to be found within the core radius of the cluster. Additionally, a simulation by \cite{1994ApJ...431L.115S} demonstrated that blue straggler stars, formed through stellar collisions in the core of the globular cluster M3, may be ejected into the outer regions due to the recoil from these interactions. We speculate that WD01 may have also been ejected into the outer regions or formed there as a result of similar interactions.  The binary system WD15 is situated far from the core of the cluster, which raises questions about its origin. Therefore, the origin of the binary system WD15 outside the cluster's core remains uncertain.

The massive WDs found in this cluster could lead to more WD mergers with a possible outcome of neutron star/pulsar formation. In the models of \cite{Kremer_2021}, they assume this will occur if the total mass of the colliding WDs exceeds the Chandrasekhar
limit, though it could be slightly more. As young millisecond pulsars have short lifetimes, those formed in the early times are no longer observed at present-day globular clusters. Therefore, currently observed young pulsars are likely formed by merging WDs, WD-MS collisions, along with the recycling of neutron stars through binaries \citep{2014A&A...561A..11V,2019ApJ...877..122Y}. Recently, a massive WD companion to the pulsar PSR J2140$-$2311B in the core-collapse globular cluster M30 was detected by \cite{Balakrishnan_2023}. With the presence of young massive WDs, NGC 362 is a cluster that may host the progenitors of millisecond pulsars. If the massive WDs of NGC 362 collide, it can lead to the formation of a Type Ia Supernova or a neutron star as per Figure 5 of  \cite{2023ApJ...944....6K}. If one of the merging WDs has a high magnetic field, this might result in the formation of a magnetar.

We note that the central core of this cluster is not resolved in the UVIT images, and it may harbour more massive WDs. The cluster is dynamically active with the formation of merged WDs and WD-MS binaries. Therefore, formation of a magnetar through a WD-WD merger is possible to happen in this globular cluster, and may provide the missing link in the formation pathway of FRB repeaters in a globular cluster in M81 \citep{2023ApJ...944....6K}. The massive WDs of NGC 362 are, therefore, unique testbeds that will provide important inputs to the properties of WDs found in core-collapsed globular clusters. 

\section {Summary and Conclusions}
We present a multiwavelength study of 17 WDs in the globular cluster NGC 362 utilizing UVIT/AstroSat and HST data sets. The following conclusions are drawn.
\begin{enumerate}
\item The UV and UV-optical CMDs were employed to select and classify the WDs. Out of twenty-three WDs, seventeen are found to be isolated.
\item Through SED analysis, we determined the fundamental parameters of isolated WDs. The effective temperature, radius, luminosity, and mass of WDs range between 22000 - 70000  K, 0.008 - 0.028 $R_\odot$, 0.09 - 3.0 $L_\odot$, and 0.30 - 1.13 $M_\odot$, respectively. 
\item We find that 14 WDs are single, while 3 (WD15, WD16, and WD17) are in binary systems. The binary systems consist of a WD with an M dwarf star. The effective temperature, radius, luminosity, and mass of M dwarfs companions are 3500 - 3750  K, 0.150-0.234 $R_\odot$, 0.003-0.01 $L_\odot$, and 0.14-0.24 $M_\odot$, respectively.
\item The various formation mechanisms of massive WDs and WD-MS binary systems have been discussed. The cooling age of the WDs implies that they were formed very recently ($\le$ 35 Myr). The binary WDs (WD15, WD16, and WD17) and massive (WD1 and WD15) WDs may have originated due to the dynamical processes that happened during the core-collapse of the cluster. This study provides the first detection of massive WDs in a core-collapsed globular cluster. We suggest that NGC 362 hosts stellar systems that may give rise to neutron stars, type Ia supernovae, and/or FRBs in the future. Hence, NGC 362 is an ideal test-bed to explore the formation pathways of these exotic objects in a globular cluster.

\end{enumerate}    
\bibliography{sample631}{}
\bibliographystyle{aasjournal}
\end{document}